      [1999/12/01 v1.4c Il Nuovo Cimento]
\documentclass{cimento}

\def\arcsec{$^{\prime\prime}$}

\usepackage{graphicx}  
\title{Probing the Nature of EROs through ASTRO-F/AKARI observations}
\author{A.~Mignano\from{ins:x}\ETC,
P.~Saracco\from{ins:x},
\atque
M.~Longhetti\from{ins:x}}
\instlist{\inst{ins:x} 
INAF - Osservatorio Astronomico di Brera, Milano - Italy}
\PACSes{\PACSit{98.52.Eh}{Elliptical galaxies}}

\begin{document}

\maketitle

\begin{abstract}
We present a preliminary analysis of ASTRO-F data of a complete sample of
$\sim$ 150 EROs (R-K$>$5) down to K$_{Vega}<$19, for which reliable photometric
redshifts are available, in the range 0.8$<$z$<$2,
selected over two fields (S7 and S2) of the MUNICS survey. The area covered is
about 420 arcmin$^2$. We have imaged this area with AKARI telescope in N3
(3.4 $\mu$m), N60 (65 $\mu$m) and WL (150 $\mu$m) down to 12 $\mu$Jy in the N3 filter, in
order to detect the rest frame H or K-band emission, thus providing an
excellent sampling of the SED of our EROs. From a first analysis we have
an identification rate of $\sim$ 63\% in the N3 filter over the S7 field.
These data allow us to distinguish starburst from passive early
type phenomena, to meseaure the SFR of the starburst component and 
to constrain the mass assembly of early type galaxies.

\end{abstract}

\section{Introduction}
While the general build-up of cosmic structures seems to be well described
by hierarchical models of galaxy formation ($\Lambda$CDM), the assembly of the
baryonic mass on galactic scale still represents a weak point of the
galaxy formation models. One of these difficulties consists in explaining
the large population of Extremely Red Objects (R-K$>$5) EROs. This
population is a mixture of early type galaxies (ETG) and dusty star forming 
galaxies (SFG). 
Hierarchical models should reproduce the abundance of massive red galaxies
at 1$<z<$2 and the balance among the number of early type and obscured
star-forming galaxies \cite{ref:Nagamine}.
Mid/Far-IR observations can play a key role in the study of this pupulation: 
such data can allow to disantangle between the early type and star forming 
phenomena and, in particular, mid-IR data allow to assess a more reliable estimate of
stellar masses of EROs, as they allow to sample the rest-frame near-IR emission.
Since the mid-IR observations are usually not available for large samples
of EROs, the rest frame H-K emission is extrapolated from the observed
optical emission (partially affected by extinction) which is dominated by
young stars ($<$0.5-1Gyr), whose contribution to the near-IR emission (dominated
by older stars) is negligible.
As the EROs are mainly composed of old stars, the
reliability of stellar masses estimates is extremely uncertain.
The way to provide the models with robust observational constraints passes
through mid-IR observations of a large ($>$100), bright (K$<$19) and complete
sample of EROs.

\begin{figure}
\centering
\includegraphics[width = 6 cm, height = 5cm]{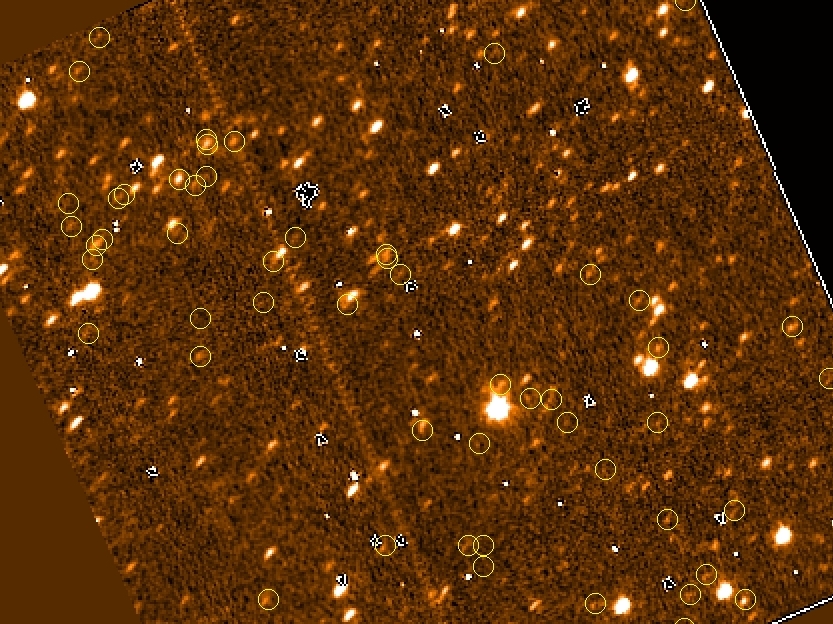}    
\includegraphics[width = 7 cm, height = 5.3cm]{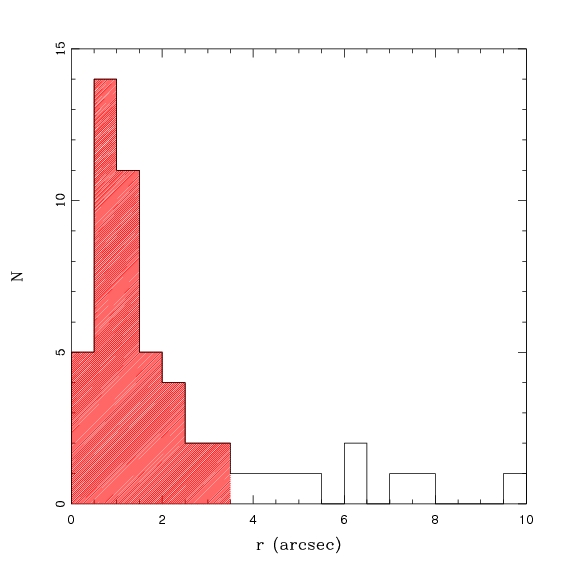}
\label{detection}
\caption{({\it left panel}) The S7F5 field imaged with N3 filter (IRC camera). The yellow circle
represent the EROs selected from the MUNICS catalogue.
({\it right panel}) N3 counterpart distribution vs. r. Red histogram shows the counterparts included
in the identification sample.}
\end{figure}

\section{The EROs Sample. Optical and IR Data}

We selected a complete sample of $\sim$150 EROs (R-K$>$5) at K$<$19 over two well
separated fields (S7 and S2) of the MUNICS survey \cite{ref:Drory}. This
area ($\sim$ 420 arcmin$^2$) and the redshift range (0.8$<$z$<$2) covered by our
sample imply a volume larger than 106 Mpc$^3$. MUNICS provides multiband
information (B,V,R,I,J,K)  down to Vega limiting magnitudes 25, 24, 23, 21, 19.5
 and photometric redshifts \cite{ref:Drory}.
We have imaged this area with the AKARI telescope in N3 (3.4 $\mu$m), N60
(65 $\mu$m) and WL (150 $\mu$m) down to 12 $\mu$Jy in the N3 filter.
This value is shallower by a factor $\sim$ 2 respect to what expected on 
the basis of AKARI Exposure Time Calculator, calculated in order to
detect in this passband all the EROs present in the field.
In this work we present the preliminary analysis made on the N3 images of
S7 field (see Fig. \ref{detection}, left panel), where 65 EROs are present.
We searched for the nearest  mid-IR object to identify the N3 counterparts of our selected EROs.
Fig. \ref{detection} (right panel) shows the mid-IR optical distance
distribution of the N3 counterparts. As it can be seen, most of identifications
are found within 3.5\arcsec which is consistent with the combined astrometric errors. 
Beyond this distance the distribution is flat as
expected for spurious identifications. Thus, we identify 41/65 EROs with a
success rate of 63\% and a contamination rate of 7\%. 
We have calculated the contamination rate,
producing several mock catalogs by randomly shifting
the EROs positions and searching for the nearest AKARI identification
within 3.5\arcsec.

\begin{figure}
\centering
\includegraphics[width = 6 cm, height = 6cm]{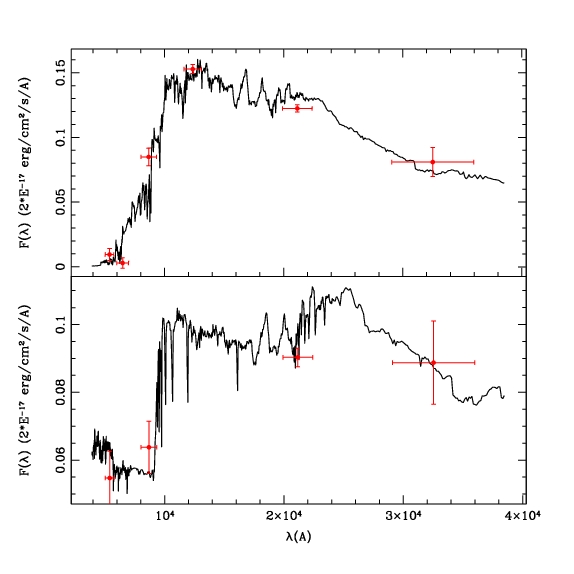}   
\includegraphics[width = 6 cm, height = 6cm]{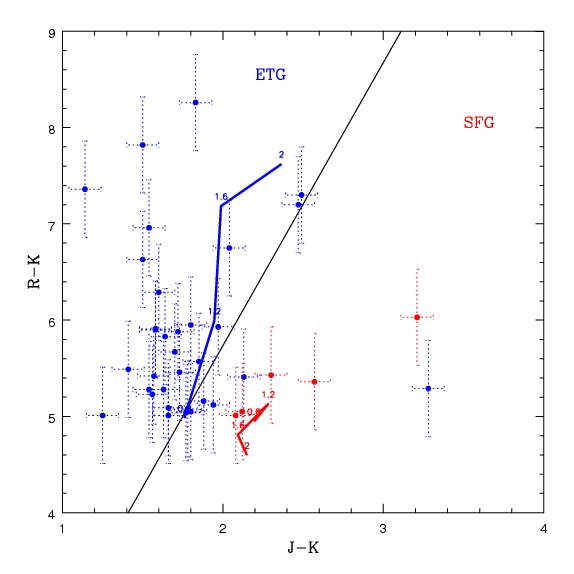}     

\label{spectra}    
\caption{({\it left panel}) Examples of fitting with SEDs of SFG and ETG.
({\it right panel}) J-K vs. R-K colours for the 41 identified EROs. 
Solid line divides the two different classes of ETG (blu points) 
and SFG (red points)
}
\end{figure}

\section{Sample Composition}
We performed a SED fitting at the fixed photometric redshifts  
running the Hyperz \cite{ref:Bolzonella} routine to assess the spectral
type of our 41 identified EROs. We used a template with an exponentially
declining SFR with e-folding time ($\tau$=0.1Gyr) and A$_{\rm V}$ $<$0.2 to describe the
early type galaxies and a constant star formation model with A${\rm V}$ $<$2.0 to describe
star forming galaxies. 
The templates have been built with the Charlot \& Bruzual (CB 2007 \cite{ref:CB07})
code, assuming the Chabrier initial mass function and solar metallicity.
The spectral type is assigned by the fit with the lowest
reduced $\chi^2$ value. 35 objects (85\% of the whole sample) were classified as
ETGs (Early Type Galaxies), while the remaining 6 (15\%) as SFGs (Star Forming
Galaxies). The choice of A$_{\rm V}<0.2$ for the ETGs implies that these galaxies
can be reddened only by old ages (and not by dust), preventing
any degeneration.
In Fig. \ref{spectra} (left panel) we show an example of ETG and SFG
galaxy, respectively . We stress that part of the SFG component can be misclassified by
the routine, thus explaining their low number.
In Fig. \ref{spectra} (right panel) we show the J-K vs. R-K colours of our sample. The solid line
R-K =  $0.34\cdot$(J-K) +0.05 represent the border between passively evolving and star forming galaxies
\cite{ref:Mannucci}. Also plotted the evolutionary tracks for ETGs (blu
line) and SFGs (red line) in the redshift range 0.8$<$z$<$2.   Our
classification broadly agrees with the Mannucci plot, but as we
may notice, a considerable number of galaxies populate the area very close to 
the partition line. In particular while none of the SFG we classified 
fall in the ETG zone, some of the ETGs fall in the SFG zone,
confirming the hypothesis of an overestimate
of the ETG class of objects.

\section{Stellar Mass Density}

We calculated the Stellar Mass for each galaxy from the best fitting model,
following the equation:
\begin{equation} 
M = \frac{2.0\cdot10^{-17}\cdot b\cdot4\cdot\pi\cdot D_L^2\cdot M_{model}}{3.826\cdot10^{33}}M_{\odot}
\end{equation}

where D$_L$ is the luminosity distance, b is a normalization factor as output of
hyperz and M$_{model}$ is the integral of the SFR of the model at the age fitted by the routine. 
The mass calculated result to be an overestimate of the real mass locked in stars,
as we do not take into account the so-called {\it return fraction}, i.e. the gas processed 
into stars and returned to the ISM.

\begin{figure}[h!]
\centering
\includegraphics[width = 6 cm, height = 6cm]{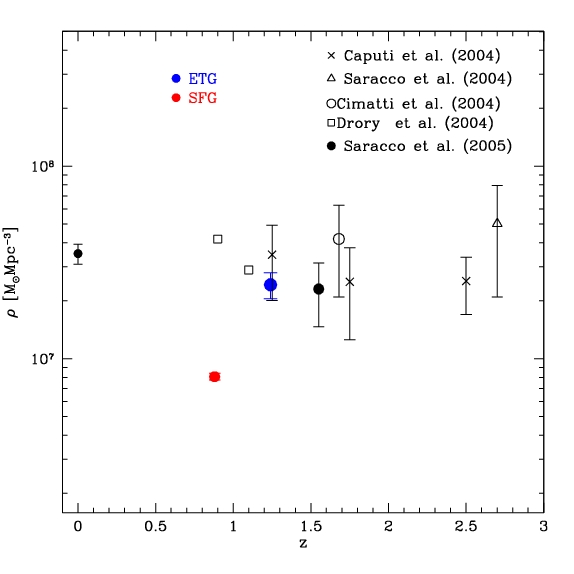}     
\label{Densityplot}
\caption{Stellar Mass Density vs. z for ETGs (blue points) and for SFGs (red points) with other literature values.}
\end{figure}

In Fig \ref{Densityplot} we show the evolution in $z$ of the stellar mass
density compared with other values in the literature.
We obtained  a stellar mass density $\rho_{*}=(2.4 \pm 0.4)\cdot 10^7M_{\odot}{\rm Mpc}^3$ 
for the ETG class. 
If we compare this value with the one obtained locally  
integrating the K--band luminosity function of the early type galaxies
\cite{ref:saracco2005},
 we can conclude that 
{\bf there is no evolution till z $\sim$ 1.3 in the
stellar mass density of the ETG population, constraining the assembly of
such objects to $z>$2.}

We notice that our results are consistent with other literature values, although
the authors we compare with do not refer to the same selection criteria. 
In particular Cimatti et al. \cite{ref:cimatti} (open circle) take into account 
the density of spheroidal galaxies independently on their masses, 
Saracco et~al. \cite{ref:saracco2004} and
Caputi et~al. \cite{ref:caputi} (open triangles and crosses respectively) 
represent the density of M$>10^{11}M_{\odot}$ evolved galaxy
candidates and Drory et~al.  \cite{ref:Drory} (open squares) show
the upper limit to the number density of galaxies 
with M$>2\cdot 10^{11}M_{\odot}$.
In fact there is no evidence of significant evolution in the 
stellar mass density of the most massive galaxies, in contrast with 
hierarchical models (e.g. \cite{ref:Delucia}) which predict that most ($\sim80$\%) of the high mass ETGs 
should complete their growth at $z<1$.

\end{document}